\documentclass[12pt]{article}
\usepackage{graphicx,amssymb,amsmath,empheq,amsthm}
\usepackage{authblk}
\usepackage{comment}
\usepackage{braket}
\usepackage{subfig} 
\usepackage{soul,}
\usepackage{hyperref}
\hypersetup{colorlinks = true, linkcolor=black, citecolor=black, urlcolor=blue}
\usepackage{url}
\usepackage{enumitem}
\usepackage[nosort]{cite}
\usepackage{amsmath,bm}

\theoremstyle{plain}

\usepackage{tikz}
\usepackage{tikz-cd}
\usetikzlibrary{arrows}
\usetikzlibrary{intersections}
\usetikzlibrary{shapes.geometric}
\usetikzlibrary{decorations.pathmorphing, patterns,shapes}
\usetikzlibrary{decorations.markings}

\tikzset{
	mid arrow/.style={postaction={decorate,decoration={
				markings,
				mark=at position .575 with {\arrow{stealth}}
	}}},
	near arrow/.style={postaction={decorate,decoration={
				markings,
				mark=at position .275 with {\arrow{stealth}}
	}}},
	far arrow/.style={postaction={decorate,decoration={
				markings,
				mark=at position .800 with {\arrow{stealth}}
	}}},
	snake arrow/.style={fixed point arithmetic, decorate, decoration={snake,amplitude=2pt, segment length=11pt},postaction={decoration={markings,mark=at position 0.625 with {\arrow{stealth}}},decorate}},
}

\tikzset{
  baseline = -0.5ex,
  wavy/.style = {
    thick,
    decorate,
    decoration={snake,amplitude=2pt,segment length=5pt}},
  swavy/.style = {
    thick,
    decorate,
    decoration={snake,amplitude=1.2pt,segment length=3pt}},
  sdot/.style = {
    circle,
    draw=none,
    fill=black,
    minimum size=2.5pt,
    inner sep=0pt},
  bdot/.style = {
    circle,
    draw=none,
    fill=black,
    minimum size=4pt,
    inner sep=0pt},
  svertex/.style = {
    circle,
    draw=black,
    thick,
    fill=lightgray,
    minimum size=8pt,
    inner sep=1pt},
  mmvertex/.style = {
    circle,
    draw=black,
    thick,
    fill=lightgray,
    minimum size=10pt,
    inner sep=1pt},
  mvertex/.style = {
    circle,
    draw=black,
    thick,
    fill=lightgray,
    minimum size=12pt,
    inner sep=1pt},
  bvertex/.style = {
    circle,
    draw=black,
    thick,
    fill=lightgray,
    minimum size=16pt},
  bbvertex/.style = {
    circle,
    draw=black,
    thick,
    fill=lightgray,
    minimum size=20pt}
 }

\title{Refining the Understanding of Operator Size Dynamics in Open Quantum Systems}

\author[1]{Haolin Jiang}
\author[1,2,3,4]{Pengfei Zhang\thanks{PengfeiZhang.physics@gmail.com}}

\affil[1]{\normalsize \it Department of Physics, Fudan University, Shanghai, 200438, China}
\affil[2]{\normalsize \it State Key Laboratory of Surface Physics, Fudan University, Shanghai, 200438, China}
\affil[3]{\normalsize \it Hefei National Laboratory, Hefei 230088, China}
\affil[4]{\normalsize \it Shanghai Qi Zhi Institute, AI Tower, Xuhui District, Shanghai 200232, China}

\date{\today}

\begin{document}
\maketitle

\begin{abstract}
Information scrambling refers to the phenomenon in which local quantum information in a many-body system becomes dispersed throughout the entire system under unitary evolution. It has been extensively studied in closed quantum systems, where it is quantified by operator size growth, revealing deep connections between condensed matter physics, high-energy physics, and quantum information. However, when extending the study of operator size dynamics to open quantum systems, two different definitions of operator size distributions emerge. These definitions are based on different treatments of the bath. In this work, we aim to establish a unified picture for operator size dynamics in open quantum systems, using the solvable Brownian SYK models at generic system size. In particular, we provide the conditions under which the signature of scrambling transition—discovered using one particular definition—appears in operator size dynamics under the other definition. Additionally, we extend previous studies by exploring finite-size effects that are not captured by the scramblon theory. Our results provide a refined understanding of operator size dynamics in open quantum systems. 
\end{abstract}

\bigskip

\section{Introduction}

Under unitary evolution, the initial quantum information in closed many-body systems is fully preserved. Nevertheless, it spreads throughout the entire system in a highly nontrivial manner, hiding quantum information in nonlocal degrees of freedom. This process, known as information scrambling \cite{Hayden:2007cs, Sekino:2008he, Shenker:2014cwa, kitaev2014talk}, has attracted extensive investigation from the perspectives of condensed matter physics, high-energy physics, and quantum information \cite{Qi:2018ogs, Swingle:2018ekw}, including experimental studies \cite{2015Natur,Li:2016xhw,Garttner:2016mqj,2019Sci...364..260B,2019arXiv190206628S,2019Natur,2020PhRvL.124x0505J,Blok:2020may,2021PhRvA.104a2402D,2021PhRvA.104f2406D,Mi:2021gdf,Cotler:2022fin,2022PhRvA.105e2232S,Liang:2024bqn,Li:2024pta}.  In the Heisenberg picture, the information scrambling can be quantified by the ``size'' of local operators, which has been studied in various contexts \cite{Roberts:2014isa,Nahum:2017yvy,vonKeyserlingk:2017dyr,Khemani:2017nda,Hunter-Jones:2018otn,Roberts:2018mnp,Chen:2019klo,PhysRevLett.122.216601,Chen:2020bmq, Lucas:2020pgj,Yin:2020oze,Zhou_Brownian_2019,Qi:2018bje,Dias:2021ncd,2021PhRvR...3c2057W,gu2022two,Yao:2024hua,Zhang:2022fma,Zhang:2023vpm,Liu:2023lyu,Tian-GangZhou:2024vxm,Xu:2024owa}. To be concrete, let us focus on systems consisting of $N$ Majorana fermions $\chi_i$ with $i\in\{1,2,...,N\}$, which satisfy the canonical anticommutation relation $\{\chi_i,\chi_j\}=2\delta_{ij}$. We expand a generic operator $O(t)$ in the basis of Majorana strings, yielding
\begin{equation}\label{eqn:closed_def}
  O(t)=\sum_n\sum_{j_1<j_2...<j_n}c_{j_1j_2...j_n}(t)~i^{\frac{n(n-1)}{2}}\chi_{j_1}...\chi_{j_n}.
  \end{equation}
  Here, $i^{\frac{n(n-1)}{2}}$ is introduced such that $i^{\frac{n(n-1)}{2}}\chi_{j_1}...\chi_{j_n}$ is Hermitian. The coefficients $c_{j_1j_2...j_n}(t)$ are the operator wavefunction, which encodes the complete information about the operator's dynamics. To define the operator size for $O(t)$, we first assign a ``size'' to each basis operator. Following the standard convention \cite{Roberts:2018mnp}, this corresponds to counting the number of nontrivial Pauli operators: $\text{Size}\Big(i^{\frac{n(n-1)}{2}}\chi_{j_1}...\chi_{j_n}\Big)=n$. Then, the operator size distribution $P(n,t)$ of a generic operator $O(t)$ is defined as the total weight of basis operators with size $n$, given by
  \begin{equation}
  P(n,t) = \sum_{j_1<j_2...<j_n} \left| c_{j_1 j_2 ... j_n}(t) \right|^2.
  \end{equation}
  It is straightforward to check that the operator size distribution satisfies the normalization condition $\sum_{n=0}^N P(n,t)=2^{-\frac{N}{2}}\text{tr}[O^\dagger O]=1$. Here, we choose a proper normalization for the operator $O$. This justifies viewing the operator size distribution as a classical distribution function. The study of operator size dynamics in many-body systems aims to extract such a classical distribution from complex quantum dynamics. We can also compute its moments. In particular, the first moment, the expectation of average size, is related to the out-of-time-order correlator (OTOC) \cite{kitaev2014talk,1969JETP...28.1200L} 
  \begin{equation}
  \overline{n}(t)=\sum_{n} n~P(n,t)=\frac{N}{2}\Big(1-\sum_j\langle O(t)\chi_jO(t)\chi_j\rangle\Big),
  \end{equation}
  where we assume that the operator $O$ is fermionic. $\langle ... \rangle$ is defined over the infinite temperature ensemble. Generalization to finite temperatures has also been discussed in \cite{Qi:2018bje,gu2022two,Zhang:2022fma}. 

  \textbf{Operator Size in Open Systems.--} Meanwhile, there is growing interest in understanding the novel quantum dynamics of many-body systems under non-unitary evolution. One important class of such systems, with high experimental relevance, is open quantum systems. In particular, the information scrambling in open quantum systems has been explored in a series of works \cite{Chen:2017dbb, Zhang:2019fcy, Almheiri:2019jqq, Zhang:2023xrr, Weinstein:2022yce, PhysRevResearch.5.033085, Bhattacharya:2022gbz, Schuster:2022bot, Bhattacharjee:2022lzy, Bhattacharjee:2023uwx,Zhang:2024vsa,PhysRevD.110.086010,Garcia-Garcia:2024mdv}; however, it has been studied under two different definitions, based on different treatments of the bath. We consider a microscopic model of the entire system, described by $H=H_s+H_b+H_{sb}$, where $H_s/H_b$ denotes the Hamiltonian of the system/bath and $H_{sb}$ describes the system-bath coupling. To be concrete, we assume that the bath contains $M$ Majorana fermions $\psi_a$, with $a\in\{1,2,...,M\}$ and $M\gg N$. Two different treatments are 
  \begin{enumerate}[label= $\circ$]
  \item Def. I:  We evolve an operator using the Heisenberg equation with the total Hamiltonian. Similar to the closed-system case, we have
  \begin{equation}\label{eqn:closed_def}
  O_{\text{I}}(t)=\sum_{m=0}^M\sum_{n=0}^N\sum_{j_1<j_2...<j_n}\sum_{a_1<a_2...<a_m}c_{j_1j_2...j_na_1a_2...a_m}(t)~i^{\frac{(n+m)(n+m-1)}{2}}\chi_{j_1}...\chi_{j_n}\psi_{a_1}...\psi_{a_m}.
  \end{equation}
  In \cite{Zhang:2023xrr}, the authors introduce the size of a basis operator by counting the number of system fermions: $\text{Size}\Big(i^{\frac{(n+m)(n+m-1)}{2}}\chi_{j_1}...\chi_{j_n}\psi_{a_1}...\psi_{a_m}\Big)=n$. This leads to the first definition of the operator size distribution
  \begin{equation}\label{eqn:opensize1}
  P_{\text{I}}(n,t) = \sum_m\sum_{j_1<j_2...<j_n}\sum_{a_1<a_2...<a_m} \left| c_{j_1 \dots j_n a_1 \dots a_m}(t) \right|^2.
  \end{equation}
  It is straightforward to check that, under this definition, we have $\sum_n  P_{\text{I}}(n,t) =1$ and 
  \begin{equation}\label{eqn:sumdef1}
  \overline{n}_{\text{I}}(t)=\sum_{n} n~P_{\text{I}}(n,t)=\frac{N}{2}\Big(1-\sum_j\langle O_\text{I}(t)\chi_jO_\text{I}(t)\chi_j\rangle_{sb}\Big).
  \end{equation}
  Here, the subscript $sb$ implies that the expectation is performed in the full Hilbert space. 

  \item Def. II: We perform the Born-Markov approximation and focus on the system degrees of freedom. The evolution of the reduced density matrix is governed by the Lindblad master equation. Equivalently, the operators are evolved under 
  \begin{equation}
  \frac{d}{dt}O_{\text{II}}=-i[H_s,O_{\text{II}}]+\sum_i \left(L_i^\dagger O_{\text{II}}L_i-\frac{1}{2}\{L_i^\dagger L_i,O_{\text{II}}\}\right),
  \end{equation}
  where $L_i$ is the Lindblad jump operator. We then decompose $O_{\text{II}}(t)$ as in closed systems:
  \begin{equation}
  O_{\text{II}}(t)=\sum_n\sum_{j_1<j_2...<j_n}\tilde{c}_{j_1j_2...j_n}(t)~i^{\frac{n(n-1)}{2}}\chi_{j_1}...\chi_{j_n}.
  \end{equation}
  This leads to the second definition of the operator size distribution \cite{Schuster:2022bot}
  \begin{equation}\label{eqn:opensize2}
  P_{\text{II}}(n,t) = \sum_{j_1<j_2...<j_n} \left| \tilde{c}_{j_1 j_2 ... j_n}(t) \right|^2,
  \end{equation}
  which has a time-dependent normalization $\mathcal{N}(t)=\sum_n P_{\text{II}}(n,t)=\langle O_{\text{II}}(t)O_{\text{II}}(t)\rangle_s$. Here, the subscript $s$ highlights that $O_{\text{II}}(t)$ is only supported on the system. We also have
  \begin{equation}
  \overline{n}_{\text{II}}(t)=\sum_{n} n~P_\text{II}(n,t)=\frac{N}{2}\Big(1-\sum_j\langle O_\text{II}(t)\chi_jO_\text{II}(t)\chi_j\rangle_{s}\Big).
  \end{equation}
  Note that this OTOC does not match the OTOC of the total microscopic Hamiltonian.

  \end{enumerate}

  These two definitions are obviously inequivalent and can lead to different physical phenomena. For example, Definition I results in novel dynamical transitions in operator size dynamics, where the system transitions from a scrambling phase, where the operator size grows to an $O(N)$ value, to a dissipative phase, where the operator size decays to zero \cite{Weinstein:2022yce, Zhang:2023xrr}. In contrast, under Definition II, the operator size always decays in the long-time limit \cite{Schuster:2022bot} \footnote{Note that in Ref. \cite{Schuster:2022bot} the expectation of operator size is defined as $\overline{n}_{\text{II}}(t)/\mathcal{N}(t)$ }. Therefore, it is natural to clarify the relationship between the two definitions and investigate the conditions under which a signature of the scrambling transition can be observed in Definition II. It is also of particular interest to understand the behavior of operator dynamics in systems with a finite number of qubits, extending previous analyses performed in the large-$N$ limit \cite{Zhang:2023xrr}. In this work, we answer these questions using the solvable Brownian Sachdev-Ye-Kitaev (SYK) models \cite{SY,Kit.KITP.2,maldacena2016remarks,Saad:2018bqo,Sunderhauf:2019djv}. Our results refine the understanding of the interplay between information scrambling and dissipation.

\section{Model \& Setup}
\begin{figure}[tb]
    \centering
    \includegraphics[width=0.98\linewidth]{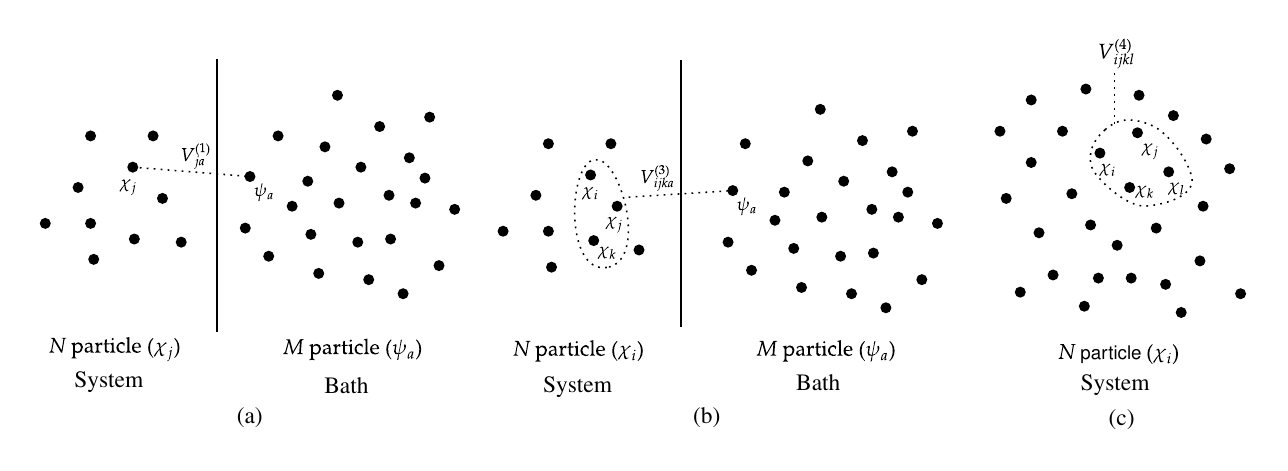}
    \caption{Schematics of various terms in the Brownian SYK model, including (a) direct hopping, (b) interactions involving three system fermions and one bath fermion, and (c) four-fermion interactions within the system. }
    \label{fig:schemticas}
  \end{figure}

We model both the system Hamiltonian $H_s$ and the system-bath coupling $H_{sb}$ using Brownian SYK models, which allow for an efficient representation in terms of effective spin degrees of freedom \cite{Sunderhauf:2019djv, Yao:2024hua, Xu:2024owa}. The specific choice for the bath Hamiltonian, $H_b$, will be irrelevant. We compare two different models, referred to as Model $A$ and Model $B$:
\begin{equation}
\begin{aligned}
    H_{A}(t) &= H_1(t) + H_3(t) = \sum_{j,a}V_{ja}^{(1)}(t)i\chi_j\psi_a 
    + \sum_{i<j<k,a}V_{ijka}^{(3)}(t) \chi_i\chi_j\chi_k\psi_a,\\
    H_{B}(t) &= H_1(t) + H_4(t) = \sum_{j,a}V_{ja}^{(1)}(t)i\chi_j\psi_a 
    + \sum_{i<j<k<l}V_{ijkl}^{(4)}(t) \chi_i\chi_j\chi_k\chi_l.
\end{aligned}
\end{equation}
A schematic of these models are presented in Figure \ref{fig:schemticas}. Model $A$ includes only two different types of system-bath coupling terms, distinguished by the number of system fermions involved. Model $B$ contains both direct hopping between the system and bath, as well as a four-fermion interaction within the system. All coupling constants are random Brownian variables, independently drawn from Gaussian distributions at each time step, satisfying
\begin{equation}\label{eqn:randomBrownianvar}
    \begin{aligned}
    \overline{V_{ja}^{(1)}(t_1)V_{ja}^{(1)}(t_2)} &= \dfrac{V_{1}}{M}\delta(t_1-t_2),\\
    \overline{V_{ijka}^{(3)}(t_1)V_{ijka}^{(3)}(t_2)} &= \dfrac{2V_{3}}{N^2M}\delta(t_1-t_2), \\
    \overline{V_{ijkl}^{(4)}(t_1)V_{ijkl}^{(4)}(t_2)} &= \dfrac{3!V_{4}}{N^3}\delta(t_1-t_2).
    \end{aligned}
\end{equation}
We focus on the limit $M\rightarrow \infty$, while keeping $N$ generic. This is essential for interpreting the $\psi_a$ fermions as a bath. The prefactors in \eqref{eqn:randomBrownianvar} are chosen to ensure a well-defined large-$M$ limit. 

    \begin{figure}[tb]
    \centering
    \includegraphics[width=0.9\linewidth]{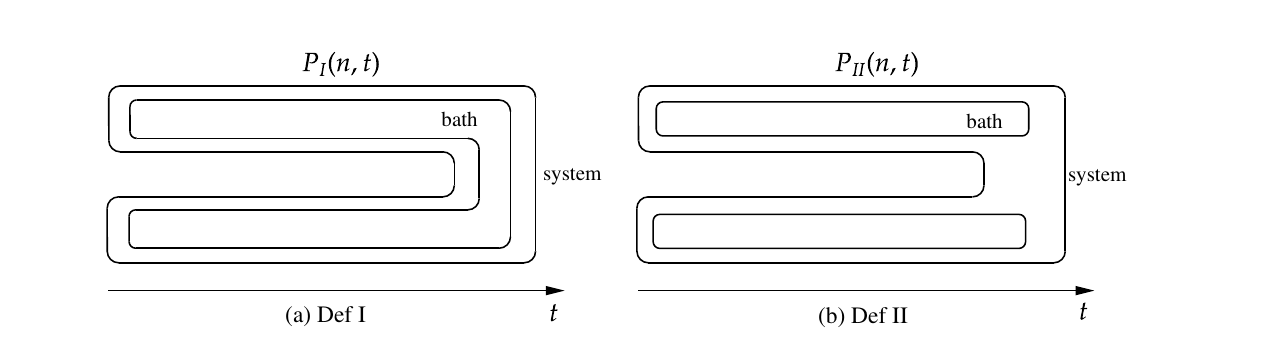}
    \caption{An illustration of the evolution contours for two different definitions, differing only in the boundary condition of the bath. Notably, only Definition I aligns with the doubled Keldysh contour used for computing the traditional OTOC in the total system \cite{aleiner2016microscopic,Gu:2018jsv,Zhang:2020jhn}. }
    \label{fig:contour}
  \end{figure}

To study operator growth, we first express both definitions of the operator size distribution using the operator-state mapping \cite{Qi:2018bje}. This involves introducing auxiliary fermion degrees of freedom $\chi^2_j$ and $\psi^2_a$. For consistency, we add a superscript $1$ to denote the original physical fermion modes as $\chi^1_j$ and $\psi^1_a$. We prepare the maximally entangled state $|12\rangle_{sb}=|12\rangle_{s}\otimes |12\rangle_{b}$, which is defined by requiring  
\begin{equation}
(\chi^1_j-i\chi^2_j)|12\rangle_{s}=0,\ \ \ \ \ \ (\psi^1_a-i\psi^2_a)|12\rangle_{b}=0.
\end{equation}
Now, we express both $P_{\text{I/II}}(n,t)$ using the operator-state mapping:
\begin{enumerate}[label= $\circ$]
\item We first focus on Definition I. Using $|12\rangle_{sb}$, we can map a generic operator $O_{\text{I}}(t)$ (which only operates in the physical Hilbert space) to a state in the doubled Hilbert space as $|O_{\text{I}}(t)\rangle\equiv O_{\text{I}}(t)|12\rangle_{sb}$, which is generally unnormalized. We can also express the Heisenberg evolution as 
\begin{equation}
|O_{\text{I}}(t)\rangle=\exp\left(-i\left(H[\chi^1_j,\psi^1_a]-H[i\chi^2_j,i\psi^2_a]\right)t\right)|O\rangle \equiv e^{-i(H^1-\tilde{H}^2)t}|O\rangle.
\end{equation}
Here, the time-ordering of the Hamiltonian evolution is kept implicit and we introduced $H^{1}$ and $\tilde{H}^2$ for conciseness. We can further express the operator size distribution as 
\begin{equation}
P_{\text{I}}(n,t)=\sum_m\sum_{j_1<...<j_n}\sum_{a_1<...<a_m}|\langle \chi_{j_1}...\chi_{j_n} \psi_{a_1}...\psi_{a_m}|O_{\text{I}}(t)\rangle|^2=\langle O_{\text{I}}(t) | \mathcal{P}_n|O_{\text{I}}(t)\rangle.
\end{equation}
Here, $\mathcal{P}_n=\sum_{j_1<...<j_n} |\chi_{j_1}...\chi_{j_n}\rangle_s\ _s\langle \chi_{j_1}...\chi_{j_n}|$ is a superoperator that projects the doubled Hilbert space onto the subspace with a fixed operator size $n$. It acts non-trivially only on the system part. Now, we apply the operator-state mapping once more to express the superoperator as a state involving four copies of fermion modes, $\chi^\alpha_j$ and $\psi^\alpha_a$, labeled by four branches of the evolution contour, $\alpha\in\{1,2,3,4\}$. Finally, this leads to \cite{Yao:2024hua}
\begin{equation}\label{eqn:PI}
P_{\text{I}}(n,t)=\ _s(\mathcal{P}_n|\otimes\ _b(14,23|e^{-i(H^1-\tilde{H}^2+H^3-\tilde{H}^4)t}|O),
\end{equation}
where $|\mathcal{P}_n)_s$ is the state that corresponds to the super-operator $\mathcal{P}_n$ and we have introduced $|O)\equiv |O\rangle_{12}\otimes|O\rangle_{34}$. $|14,23)_b=|14\rangle_{b}\otimes |23\rangle_{b}$ originates from the identity super-operator on the bath, which effectively connects branches $14$ and $23$. The overall normalization can always be fixed by requiring $\sum_n P(n,t)=1$. 

\item  Now, we turn to Definition II, where the evolution is governed by the Lindblad master equation. In general, this leads to the relation that 
\begin{equation}
O_{\text{II}}(t)\approx\text{tr}_b[O_{\text{I}}(t)\rho_b].
\end{equation}
Here, $\rho_b$ is the density matrix of the bath, which is approximated as time-independent under the Born-Markov approximation. In the Brownian SYK models, the Born approximation holds exactly in the large-$M$ limit \cite{Chen:2017dbb}, while the Markov approximation is justified by the Brownian nature of the couplings. Consequently, the relation becomes exact. Furthermore, since the Hamiltonian is time-dependent, $\rho_b$ should be fixed as the infinite-temperature ensemble. Therefore, we find $O_{\text{II}}(t)\approx\text{tr}_b[O_{\text{I}}(t)]/2^{N/2}$, or
\begin{equation}\label{eqn:partialtrace}
|O_{\text{II}}(t)\rangle=\ _b\langle 12| O_{\text{I}}(t)\rangle=\ _b\langle 12| e^{-i(H^1-\tilde{H}^2)t}|O\rangle.
\end{equation}
This selects out operators that are purely supported on the system part. As a consequence, according to Definition II, the operator size distribution can be expressed as
\begin{equation}
\begin{aligned}\label{eqn:PII}
P_{\text{II}}(n,t)&=\langle O_{\text{I}}(t) | \left(\mathcal{P}_n\otimes |12\rangle_b \ _b\langle 12|\right)|O_{\text{I}}(t)\rangle\\
&=\ _s(\mathcal{P}_n|\otimes\ _b(12,34|e^{-i(H^1-\tilde{H}^2+H^3-\tilde{H}^4)t}|O).
\end{aligned}
\end{equation}
Comparing to the Definition I, the only difference is the final state of the bath. The overall normalization can also be fixed by requiring $\sum_n P(n,0)=1$. 

\end{enumerate}

A pictorial representation of the above results is presented in Figure \ref{fig:contour}. It is inspiring to notice that only the contour under Definition I aligns with the doubled Keldysh contour used for computing the traditional OTOC in the total system \cite{aleiner2016microscopic,Gu:2018jsv,Zhang:2020jhn}. This is indeed consistent with the relation in Equation \eqref{eqn:sumdef1}, where the right-hand side represents the OTOC of the total system. Such a contour may contain a soft mode named the scramblon \cite{gu2022two,Stanford:2021bhl}, which is the origin of the exponential growth of the operator size for a parametrically long time window ($\sim\ln N$). In contrast, the contour under Definition II contains no quantum coherence between branches 12 and 34. As a result, the operator size will always decay, showing a large difference compared to the OTOC of the total system. This discrepancy arises because, while the traditional Lindblad equation guarantees the prediction of observables, the operator size involves two replicas and is not a conventional observable. The consistency can be restored by deriving a generalized master equation for multiple copies of the system \cite{Zhou:2022qhe}. On the other hand, $P_{\text{II}}(n,t)$ can be of higher experimental relevance, since its measurement does not require reversing the sign of the system-bath coupling.

    \begin{figure}[tb]
    \centering
    \includegraphics[width=0.75\linewidth]{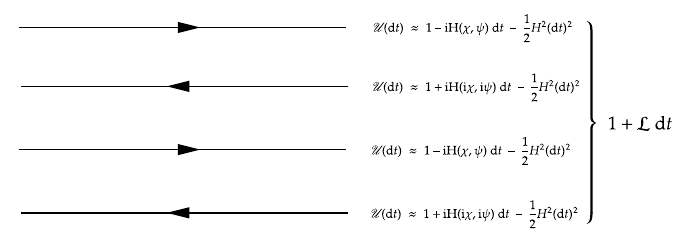}
    \caption{A sketch for the derivation of the Liouvillian $\mathcal{L}$. Here, we consider an infinitesimal time step, collect terms that are quadratic in $H dt$, and perform the disorder average. }
    \label{fig:evolution}
  \end{figure}
Having established equations \eqref{eqn:PI} and \eqref{eqn:PII}, we now perform the disorder average over the Brownian couplings given by \eqref{eqn:randomBrownianvar}. Since the couplings at different time steps are independent, we can perform the disorder average by focusing on an infinitesimally small time duration, as done in \cite{Sunderhauf:2019djv,Stanford:2021bhl, Yao:2024hua, Xu:2024owa}. In other words, we have 
\begin{equation}
\overline{e^{-i(H^1(t)-\tilde{H}^2(t)+H^3(t)-\tilde{H}^4(t))dt}}=e^{\mathcal{L}dt},
\end{equation}
The Liouvillian $\mathcal{L}$ acts on four copies of the quantum system. To derive $\mathcal{L}$, we need to expand the left-hand side to the second order in $H dt$, and note that $\delta(0)= 1/dt$ (see Figure \ref{fig:evolution} for a sketch). Here, we only present the final results for Model $A$ and Model $B$, respectively:
\begin{equation}
\mathcal{L}_A=\mathcal{L}_1+\mathcal{L}_3,\ \ \ \ \ \ \ \ \ \mathcal{L}_B=\mathcal{L}_1+\mathcal{L}_4.
\end{equation}
Here, $\mathcal{L}_i$ denotes the contribution from the Hamiltonian $H_i(t)$, which reads
\begin{equation}
\begin{aligned}
\mathcal{L}_1&=-2NV_1 - \sum_{\alpha\neq\beta} \dfrac{V_1}{M}\chi^{\alpha\beta}\psi^{\alpha\beta},\\
\mathcal{L}_3&=-\dfrac{4}{N^2}\binom{N}{3} V_3
        + \sum_{\alpha\neq\beta}
         \gamma_{\alpha,\beta}\,\dfrac{V_3}{3MN^2}
        \left[\left(\chi^{\alpha\beta}\right)^3 + (2-3N)\chi^{\alpha\beta}\right] \psi^{\alpha\beta},\\
\mathcal{L}_4&= -\dfrac{12 }{N^3}\binom{N}{4}V_4 + \sum_{\alpha \neq \beta} \gamma_{\alpha, \beta}
        \dfrac{ V_4}{4N^3}
        \left[\left(\chi^{\alpha \beta}\right)^4-\left(\chi^{\alpha \beta}\right)^2
        (-6 N+8)+3 N(N-2)\right]. 
\end{aligned}
\end{equation}
Here, $\binom{p}{q}$ is the Binomial number. We have introduced $\chi^{\alpha\beta} = \sum_{j=1}^{N}\chi_j^{\alpha}\chi_j^{\beta}$, $\psi^{\alpha\beta} = \sum_{a=1}^{M}\psi_a^{\alpha}\psi_a^{\beta}$, and phase factors due to the difference between $H$ and $-\tilde{H}$ for four fermion interactions:
\begin{equation}
    \gamma_{\alpha, \beta}= 
    \begin{cases}
        1 & (\alpha, \beta)=(1,2),(1,4),(2,3),(3,4) \\ 
        -1 & (\alpha, \beta)=(1,3),(2,4)
    \end{cases}
\end{equation}

It is straightforward to show that the Liouvillian commutes with the fermion parity operators $F_{\chi,j}=-\chi^1_j\chi^2_j\chi^3_j\chi^4_j$ for each $j$ and $F_{\psi,a}=-\psi^1_a\psi^2_a\psi^3_a\psi^4_a$ for each $a$. From now on, we focus on $O=\chi_{j_0}$, such that the initial state $|\chi_{j_0})$ is an eigenstate with eigenvalue 1 for all these parities. This reduces the number of independent components of $\chi^{\alpha\beta}$. Similar to the observations in \cite{Sunderhauf:2019djv,Stanford:2021bhl, Yao:2024hua}, the remaining components, $\chi^{\alpha\beta}$ or $\psi^{\alpha\beta}$, form two independent $SU(2)$ groups: 
\begin{equation}
\begin{aligned}
&L_x = \dfrac{1}{2i}\chi^{23}= -\dfrac{1}{2i}\chi^{14},\ \ \ \ \ \ L_y = \dfrac{1}{2i}\chi^{31}= -\dfrac{1}{2i}\chi^{24},\ \ \ \ \ \  L_z = \dfrac{1}{2i}\chi^{12}=-\dfrac{1}{2i}\chi^{34},\\
&G_x = \dfrac{1}{2i}\psi^{23}= -\dfrac{1}{2i}\psi^{14},\ \ \ \ \ \ G_y = \dfrac{1}{2i}\psi^{31}= -\dfrac{1}{2i}\psi^{24},\ \ \ \ \ \  G_z = \dfrac{1}{2i}\psi^{12}=-\dfrac{1}{2i}\psi^{34}.
\end{aligned}
\end{equation}
It has been established that $|\mathcal{P}_n)_s$ is the unnormalized eigenstate of $L_z$ with $L^2=\frac{N}{2}(\frac{N}{2}+1)$ \cite{Yao:2024hua}:
\begin{equation}
L_z|\mathcal{P}_n)_s=\left(-\frac{N}{2}+n\right)|\mathcal{P}_n)_s,\ \ \ \ \ \ \ _s(\mathcal{P}_n|\mathcal{P}_n)_s=\binom{N}{n}.
\end{equation} 
In particular, $|12,34)=|P_0)$ is the fully polarized state. This determines all matrix elements of the group generators. Since the system exhibits permutation symmetry, we can average over the initial site $j_0$. This leads to $\overline{|\chi_{j_0})}=\frac{1}{N}|\mathcal{P}_1)\otimes|12,34)_b$. Putting all ingredients together, the evolution of the system is restricted to an $N+1$-dimensional space with spin-$N/2$. 

If we make a direct analogy, we would use a single spin-$M/2$ to represent the bath. However, since we are focusing on the limit $M\rightarrow \infty$, this enlarge of the Hilbert space is not necessary. Both $G_{x,y,z}/M$ represent averages over $M$ copies of fermion operators. By the central limit theorem, their fluctuations are suppressed by $1/M$ and can be safely neglected compared to their expectations. Similar conclusion can be reached by diagrammatic analysis \cite{Chen:2017dbb,Zhang:2019fcy}. Therefore, we replace $\psi^{\alpha\beta}$, or $G_{x/y/z}$, by their expectation values, which are taken by neglecting all system degrees of freedom\footnote{One might raise concerns regarding the dynamics of the bath, which is not included in either $H_A$ and $H_B$. However, for a well-defined microscopic bath model, the scrambling time is on the order of $\ln M$, which diverges as $M\rightarrow \infty$. Therefore, the intrinsic dynamics of the bath can be safely neglected, as discussed in \cite{Chen:2017dbb}. }. This distinguishes two different definitions. For Definition I, we have 
\begin{equation}
\bm{G}\rightarrow\langle \bm{G}\rangle=\frac{\ _b(14,23|\bm{G}|12,34)_b}{\ _b(14,23|12,34)_b}=\left(\dfrac{M}{2},\dfrac{M}{2}i, -\dfrac{M}{2}\right),
\end{equation}
where $\bm{G}=(G_x,G_y,G_z)$. In contrast, for Definition II, we have
\begin{equation}
\bm{G}\rightarrow\langle \bm{G}\rangle=\frac{\ _b(12,34|\bm{G}|12,34)_b}{\ _b(12,34|12,34)_b}=\left(0,0, -\dfrac{M}{2}\right).
\end{equation}
This leads to different behaviors of the operator size dynamics under different definitions: Denoting the effective system Liouvillian after this replacement as $\mathcal{L}^{\text{I/II}}_s$, we find
\begin{equation}
P_{\text{I/II}}(n,t)=\frac{1}{N}\ _s(\mathcal{P}_n|e^{\mathcal{L}^{\text{I/II}}_s t}|\mathcal{P}_1)_s.
\end{equation}
For Definition I, this can be understood as a generalized Lindblad master equation.

Putting all ingredients together, we can derive the differential equation for the evolution of $P_{\text{I/II}}(n,t)$, which takes the form of
\begin{equation}\label{eqn:dP}
\partial_t P_{\text{I/II}}(n,t)=\frac{1}{N}\ _s(\mathcal{P}_n|\mathcal{L}^{\text{I/II}}_se^{\mathcal{L}^{\text{I/II}}_s t}|\mathcal{P}_1)_s=\sum_{\Delta n}C_{\Delta n}(n)P_{\text{I/II}}(n+\Delta n,t).
\end{equation}
Here, $C_{\Delta n}(n)$ describes the transition rate from size-$n+\Delta n$ operators to size-$n$ operators. The initial condition is $P_{\text{I/II}}(n,0)=\delta_{n1}$. Under Definition I, both $H_3$ and $H_4$ can increase the operator size. Specifically, $[H_3,\chi]=\psi\chi\chi$ ($\Delta n=-1$) and $[H_4,\chi]=\chi\chi\chi$ ($\Delta n=-2$). This is consistent with the non-zero components of $C_{\Delta n}(n)$: 
\begin{enumerate}[label= $\circ$]
\item Model A ($H_1+H_3$) under Definition I:
    \begin{equation}\label{eqn:222}
        \begin{aligned}
        C_0(n) &= -4V_1 n -\dfrac{4V_3}{3N^2}(2+4n^2 -3N-6nN+3N^2)n,\\
        C_1(n) &= 4V_1(n+1),\\
        C_{-1}(n) &= \dfrac{4V_3}{N^2}(n-1)(n-N-1)(n-N),\\
        C_{3}(n) &= \dfrac{4V_3}{3N^2}(n+3)(n+2)(n+1).
        \end{aligned}
    \end{equation}
\item Model B ($H_1+H_4$) under Definition I:
    \begin{equation}\label{eqn:111}
    \begin{aligned}
        C_0(n) &= -4V_1 n + \dfrac{4V_4n(n-N)\left(2(2+n^2) - (3+2n)N + N^2\right)}{N^3},\\
        C_1(n) &= 4V_1(n+1),\\
        C_2(n) &= -\dfrac{4V_4n(n+1)(n+2)(2+n-N)}{N^3},\\
        C_{-2}(n) &= -\dfrac{4V_4(n-2)(n-2-N)(n-1-N)(n-N)}{N^3}.
    \end{aligned}
    \end{equation}
\end{enumerate}
It is straightforward to check that $\partial_t\sum_nP_{\text{I/II}}(n,t)=0$. As a comparison, under Definition II, only $H_4$ can lead to operator size growth. This is because that $[H_3,\chi]=\psi\chi\chi$ results in a non-trivial bath operator, which is neglected when performing the partial trace in \eqref{eqn:partialtrace}. This is consistent with the results:
\begin{enumerate}[label= $\circ$]
\item Model A ($H_1+H_3$) under Definition II:
    \begin{equation}\label{eqn:modelBIIC0}
        \begin{aligned}
        C_0(n) = -4V_1 n -\dfrac{4V_3}{3N^2}(2+4n^2 -3N-6nN+3N^2)n.
        \end{aligned}
    \end{equation}
\item Model B ($H_1+H_4$) under Definition II:
    \begin{equation}
    \begin{aligned}
        C_0(n) &= -4V_1 n + \dfrac{4V_4n(n-N)\left(2(2+n^2) - (3+2n)N + N^2\right)}{N^3},\\
        C_2(n) &= -\dfrac{4V_4n(n+1)(n+2)(2+n-N)}{N^3},\\
        C_{-2}(n) &= -\dfrac{4V_4(n-2)(n-2-N)(n-1-N)(n-N)}{N^3}.
    \end{aligned}
    \end{equation}
\end{enumerate}
Importantly, operators of different sizes do not couple in Model A for arbitrary system size $N$, marking a sharp contrast with Definition I.

\section{Results}
We analyze the equation \eqref{eqn:dP} obtained in the last section, which govern the operator size dynamics in open systems under two distinct definitions. We first focus on the limit of $N\rightarrow \infty$, where analytical progress is possible. Then, we present numerical simulations for finite $N$. 

\subsection{The \texorpdfstring{$N\rightarrow \infty $}{TEXT} limit}
We consider the operator size dynamics in the limit of $N\rightarrow \infty$ with finite evolution time $t$ for both Model $A$ and Model $B$. In this regime, we only keep $C_{\Delta n}(n)$ to the zeroth order in $1/N$ and the equation are significantly simplified. 
\vspace{5pt}

\textbf{Definition I.--} This problem has been studied in \cite{Zhang:2023xrr} using the path-integral formalism. Here, we reproduce these results by analyzing the differential equation for $P_{I}(n,t)$, demonstrating the emergence of scrambling transitions in both models as $V_1$ increases. Let us consider the Model $A$ with $H=H_1+H_3$. In the large-$N$ limit, the equation \eqref{eqn:dP} reads
\begin{equation}\label{eqn:modelA}
\partial_t P_{\text{I}}(n,t)=-4(V_1+V_3) n~P_{\text{I}}(n,t)+4V_1(n+1)P_{\text{I}}(n+1,t)+4V_3 (n-1)P_{\text{I}}(n-1,t).
\end{equation}
We can solve the equation by introducing the generating function $z_{\text{I}}(\mu,t)=\sum_{n=0}^\infty e^{-\mu n}P_{\text{I}}(n,t)$. \eqref{eqn:modelA} now becomes a partial differential equation
\begin{equation}\label{eqn:partialdiff}
\partial_t z_{\text{I}}(\mu,t)=\left[4V_1(1-e^{\mu})+4V_3(1-e^{-\mu})\right]\partial_\mu z_{\text{I}}(\mu,t)\equiv A_{\text{I}}(\mu) \partial_\mu z_{\text{I}}(\mu,t),
\end{equation}
with the initial condition $z_{\text{I}}(\mu,0)=e^{-\mu}$. To proceed, we define $\frac{d\mu}{A_\text{I}(\mu)}=d x$, and observe that $z_\text{I}(\mu,t)$ should be a function of $t+x$. This leads to the result 
\begin{equation}
z_{\text{I}}(\mu,t)=\frac{\left(e^{\mu }-1\right) V_1 e^{4  (V_3-V_1)t}+ \left(V_3-e^{\mu } V_1\right)}{\left(e^{\mu }-1\right) V_3 e^{4  (V_3-V_1)t}+\left(V_3-e^{\mu } V_1\right)}.
\end{equation}
The size distribution can be obtained via a Taylor expansion in $e^{-\mu}$, which yields
\begin{equation}
P_{\text{I}}(n,t)=\begin{cases}
      1-{(V_3-V_1)e^{4(V_3-V_1)t}}/\left({V_3e^{4(V_3-V_1)t}-V_1}\right), & (n=0)\\
      {(V_3-V_1)^2e^{4(V_3-V_1)t}[V_3e^{4(V_3-V_1)t}-V_1]^{n-1}}/{[V_3e^{4(V_3-V_1)t}-V_1]^{n+1}}. & (n\geq 1)
    \end{cases} 
\end{equation}
Alternatively, we characterize the dynamics of operator size by computing its expectation value:
\begin{equation}
\overline{n}_{\text{I}}(t)=-\partial_\mu z_{\text{I}}(0,t)=e^{4(V_3-V_1)t}. 
\end{equation}
This clearly shows a scrambling transition at $V_1/V_3=1$, separating a scrambling phase for $V_3>V_1$, where the operator size grows exponentially, from a dissipative phase for $V_3<V_1$ where the operator size decays exponentially.

The above results are consistent with those obtained in \cite{Zhang:2023xrr}. As a side remark, \cite{Zhang:2023xrr} derives the evolution of $z_{\text{I}}(\mu,t)$ using the Schwinger–Dyson equation on the doubled Keldysh contour (see Figure \ref{fig:contour}(b)), rather than the linear partial differential equation \eqref{eqn:partialdiff}. Notably, their approach leads to a nonlinear equation in $z_{\text{I}}(\mu,t)$: 
\begin{equation}\label{eqn:modelAH}
\partial_t z_{\text{I}}(\mu,t)=4V_3(z_{\text{I}}(\mu,t)^2-z_{\text{I}}(\mu,t))+4V_1(1-z_{\text{I}}(\mu,t)).
\end{equation}
The scrambling transition can be determined by expanding \eqref{eqn:modelAH} to the leading order of $\mu$. It is then interesting to ask how this equation can be obtained within our approach. In fact, the equation \eqref{eqn:modelA} serves as an analog of the Schrödinger equation, while \eqref{eqn:modelAH} resembles the Heisenberg equation. To see this, we first express $z_{\text{I}}(\mu,t)$ as
\begin{equation}
\begin{aligned}
z_{\text{I}}(\mu,t)=&\frac{1}{N}\Big[\sum_ne^{-\mu n}\ _s(\mathcal{P}_n|\Big]e^{\mathcal{L}^{\text{I}}_s t}\Big(\sum_j\chi^1_j\chi^3_j\Big)e^{-\mathcal{L}^{\text{I}}_s t}|12,34)_s\\
\equiv&\frac{1}{N}(\mu|e^{\mathcal{L}^{\text{I}}_s t}\chi^{13}e^{-\mathcal{L}^{\text{I}}_s t}|12,34)_s.
\end{aligned}
\end{equation}
Here, we have inserted $e^{-\mathcal{L}^{\text{I}}_s t}$ without affecting the inner product, since $\mathcal{L}^{\text{I}}_s|12,34)=0$. Therefore, we have
\begin{equation}\label{eqn:commutator}
\partial_t z_{\text{I}}(\mu,t)=\frac{1}{N}(\mu|e^{\mathcal{L}^{\text{I}}_s t}[\mathcal{L}^{\text{I}}_s,\chi^{13}]|12,34)_s.
\end{equation}
Moreover, as established in \cite{Qi:2018bje}, $|\mu)$ is a Gaussian state and thus satisfies the Wick theorem. In the current case, this corresponds to setting $\langle \chi^{\alpha_1\beta_2}\chi^{\alpha_2\beta_2}...\chi^{\alpha_l\beta_l}\rangle=\langle \chi^{\alpha_1\beta_2}\rangle\langle \chi^{\alpha_2\beta_2}\rangle...\langle\chi^{\alpha_l\beta_l}\rangle$. By computing the commutator $[\mathcal{L}^{\text{I}}_s,\chi^{13}]$ and using the relation $\chi^{13}|12,34)=i\chi^{14}|12,34)$, we find that equation \eqref{eqn:commutator} is consistent with \eqref{eqn:modelAH}.

We can analyze Model B with $H=H_1+H_4$ using similar methods. In this case, equation \eqref{eqn:dP} takes the form:
\begin{equation}\label{eqn:modelB}
\partial_t P_{\text{I}}(n,t)=-4(V_1+V_4) n~P_{\text{I}}(n,t)+4V_1(n+1)P_{\text{I}}(n+1,t)+4V_4 (n-2)P_{\text{I}}(n-2,t).
\end{equation}
Similar to Model A, we can introduce the generating function, which satisfies
\begin{equation}\label{eqn:partialdiff}
\partial_t z_{\text{I}}(\mu,t)=\left[4V_1(1-e^{\mu})+4V_4(1-e^{-2\mu})\right]\partial_\mu z_{\text{I}}(\mu,t)\equiv B_{\text{I}}(\mu) \partial_\mu z_{\text{I}}(\mu,t).
\end{equation}
However, we are unable to obtain a closed-form solution for this equation. Instead, we compute the expectation value of the operator size directly. By expanding both sides to $O(\mu)$, we find
\begin{equation}
\partial_t \overline{n}_{\text{I}}(t)=4(2V_4-V_1)\overline{n}_{\text{I}}(t),\ \ \ \ \ \ \overline{n}_{\text{I}}(t)=e^{4(2V_4-V_1)t}.
\end{equation}
This shows a scrambling transition at $V_1/V_4=2$. Therefore, in Definition I, there is no intrinsic difference between operator growth induced by an intra-system interaction term $H_4$ and that induced by an interaction term involving bath fermions $H_3$. We can also derive the Heisenberg equation as
\begin{equation}\label{eqn:modelBH}
\partial_t z_{\text{I}}(\mu,t)=4V_4(z_{\text{I}}(\mu,t)^3-z_{\text{I}}(\mu,t))+4V_1(1-z_{\text{I}}(\mu,t)).
\end{equation}
\vspace{5pt}

\textbf{Definition II.--} We now analyze the operator size dynamics under Definition II, beginning with Model A, where the evolution of the operator size distribution is governed by
\begin{equation}\label{eqn:modelAII}
\partial_t P_{\text{II}}(n,t)= -4(V_3+V_1)n P_{\text{II}}(n,t). 
\end{equation}
This equation can be solved directly, as operators of different sizes do not couple. The result reads $P_{\text{OI}}(n,t)=e^{-4(V_1+V_3)t}\delta_{n0}$, which decays monotonically as $t$ increases. More interestingly, the decay rate increases as $V_3$ increases. This is expected, as we have observed that, in our setup, scrambling induced by system-bath interactions $H_3$ does not contribute after tracing out the bath. Therefore, we do not see any signature of scrambling transition for Model A. This holds even for finite $N$, as described by \eqref{eqn:modelBIIC0}. In addition, we can also derive the Heisenberg equation 
\begin{equation}
\partial_t z_{\text{II}}(\mu,t)=-4(V_3+V_1)z_{\text{II}}(\mu,t).
\end{equation}
Compared with equation \eqref{eqn:modelAH} for Definition I, the terms $V_3 z^2$ and $V_1$ are absent, as they originate from the $L_x$ and $L_y$ terms in $\mathcal{L}^{\text{I}}_s$. 

In contrast, the operator size can still grow in Model B due to the intra-system interaction $H_4$. In this case, equation \eqref{eqn:dP} becomes
\begin{equation}\label{eqn:modelBII}
\partial_t P_{\text{II}}(n,t)=-4(V_1+V_4) n~P_{\text{II}}(n,t)+4V_4 (n-2)P_{\text{II}}(n-2,t).
\end{equation}
Introducing the generating function $z_{\text{II}}(\mu,t)=\sum_{n=0}^\infty e^{-\mu n}P_{\text{II}}(n,t)$, we obtain the partial differential equation
\begin{equation}\label{eqn:partialdiffII}
\partial_t z_{\text{II}}(\mu,t)=\left[4V_1+4V_4(1-e^{-2\mu})\right]\partial_\mu z_{\text{II}}(\mu,t)\equiv B_{\text{II}}(\mu) \partial_\mu z_{\text{II}}(\mu,t).
\end{equation}
This equation can be solved analytically using the method outlined above. The resulting expression is:
\begin{equation}
z_{\text{II}}(\mu,t)=\frac{\sqrt{V_4+V_1}}{\sqrt{e^{8 t (V_4+V_1)} \left(e^{2 \mu } (V_4+V_1)-V_4\right)+V_4}}.
\end{equation}
This leads to the operator size distribution
\begin{equation}
P_{\text{II}}(2n+1,t)=\frac{\Gamma \left(n+\frac{1}{2}\right)}{\sqrt{\pi } \Gamma\left(n+1\right)}e^{-4(V_1+V_4)t}\left[\frac{V_4 e^{-8 t \left(V_1+V_4\right)} \left(e^{8 t \left(V_1+V_4\right)}-1\right)}{ \left(V_1+V_4\right)}\right]^n,
\end{equation}
and $P_{\text{II}}(2n,t)=0$. In particular, we can compute the expectation of the operator size as
\begin{equation}
\overline{n}_{\text{II}}(t)=-\partial_\mu z_{\text{II}}(0,t)= \frac{\left(V_1+V_4\right){}^{3/2} e^{8 t \left(V_1+V_4\right)}}{\left(V_1 e^{8 t \left(V_1+V_4\right)}+V_4\right){}^{3/2}}=1+4 t \left(2 V_4-V_1\right)+O(t^2).
\end{equation}
In the long-time limit, the operator size always decays as $e^{-4t(V_1+V_4)}$, provided that $V_1 \neq 0$. On the other hand, in the short-time regime where $V_1 t\ll 1$ and $V_4 t\ll 1$, we observe signatures of a scrambling transition: operator growth occurs for $2V_4>V_1$ while operator decay is observed for $2V_4<V_1$. Notably, the transition point for the early-time behavior precisely coincides with the scrambling transition under Definition I. To understand this coincidence better, we derive the Heisenberg equation, which reads
\begin{equation}\label{eqn:modelBHII}
\partial_t z_{\text{II}}(\mu,t)=4V_4(z_{\text{II}}(\mu,t)^3-z_{\text{II}}(\mu,t))-4V_1z_{\text{II}}(\mu,t).
\end{equation}
The only difference, comparing to \eqref{eqn:modelBH}, is the absence of a $V_1$ term, originates from the coherence between branches $(1,2)$ and $(3,4)$. Therefore, if we take the derivative of both \eqref{eqn:modelBH} and \eqref{eqn:modelBHII} with respect to $\mu$ and set $\mu=0$, we find
\begin{equation}
\begin{aligned}
\partial_t \overline{n}_{\text{I}}(t)&=8V_4\overline{n}_{\text{I}}(t)-4V_1\overline{n}_{\text{I}}(t),
\\
\partial_t \overline{n}_{\text{II}}(t)&=4V_4(3z_{\text{II}}(0,t)^2-1)\overline{n}_{\text{II}}(t)-4V_1\overline{n}_{\text{II}}(t).
\end{aligned}
\end{equation}
The difference arises because $z_{\text{I}}(0,t)=1$, while $z_{\text{II}}(0,t)$ decays over time. Nevertheless, we have $z_{\text{II}}(0,0)$. As a result, both $\overline{n}_{\text{I}}(t)$ and $\overline{n}_{\text{II}}(t)$ exhibit the same short-time behavior. This accounts for the observed signature of the scrambling transition, when the scrambling is due to the intra-system interaction.

Finally, we comment on more general models that include both intra-system interactions and system-bath couplings involving multiple system operators. An example is the Hamiltonian $H=H_1+H_3+H_4$. In this case, we expect to observe different dynamical behaviors at early times under Definitions I and II. Specifically, in Definition I, both intra-system interactions and system-bath couplings contribute to information scrambling, whereas in Definition II, only intra-system interactions lead to an increase in operator size. As a result, we expect the transition points to differ between the two definitions. 

\subsection{Finite-\texorpdfstring{$N$}{TEXT} corrections}

We then consider finite-$N$ effects. We begin with a review of related results from previous studies. In \cite{Zhang:2023xrr}, the authors applied the scramblon effective theory \cite{gu2022two,Stanford:2021bhl} to study finite-
$N$ effects in the scrambling phase of Model A under Definition I. The central assumption is that, after short-time evolution, information scrambling is dominated by a collective mode on the doubled Keldysh contour, known as the scramblon. This captures the late-time regime with fixed $e^{4(V_1-V_3) t}/N$ in the large-$N$ limit. For convenience, we reproduce the results below ($V_3>V_1$):
\begin{equation}
   \mathcal{P}_{\text{I}}(s,t)=r \delta (s)+\mathcal{P}_\text{reg}(s,t),\ \ \ \ \ 
   \mathcal{P}_\text{reg}(s,t)=\theta\left(1-\frac{2s}{1-r}\right)\frac{2 (r-1)^2 e^{\frac{2 s}{\lambda  (r+2 s-1)}}}{\lambda  (r+2 s-1)^2}.
\end{equation}
Here, we have $\lambda=e^{4(V_3-V_1)t}/C$ with $C=N(1-r)^2/2$ and $r=V_1/V_3<1$. $\mathcal{P}_{\text{I}}(s,t)$ is a continuum version of the operator size distribution, defined as $\mathcal{P}_{\text{I}}(s,t)=NP_{\text{I}}(sN,t)$, which satisfies the normalization condition $\int_0^1 ds~\mathcal{P}_{\text{I}}(s,t)=1$. This predicts that the operator size distribution approaches the combination of two delta functions in the long-time limit:
\begin{equation}
\mathcal{P}_{\text{I}}(s,\infty)=r \delta(s)+(1-r)~\delta\left(s-\frac{1-r}{2}\right),
\end{equation}  
which leads to $\overline{n}_{\text{I}}(\infty)=\frac{(1-r)^2}{2}$.

    \begin{figure}[tb]
    \centering
    \includegraphics[width=0.75\linewidth]{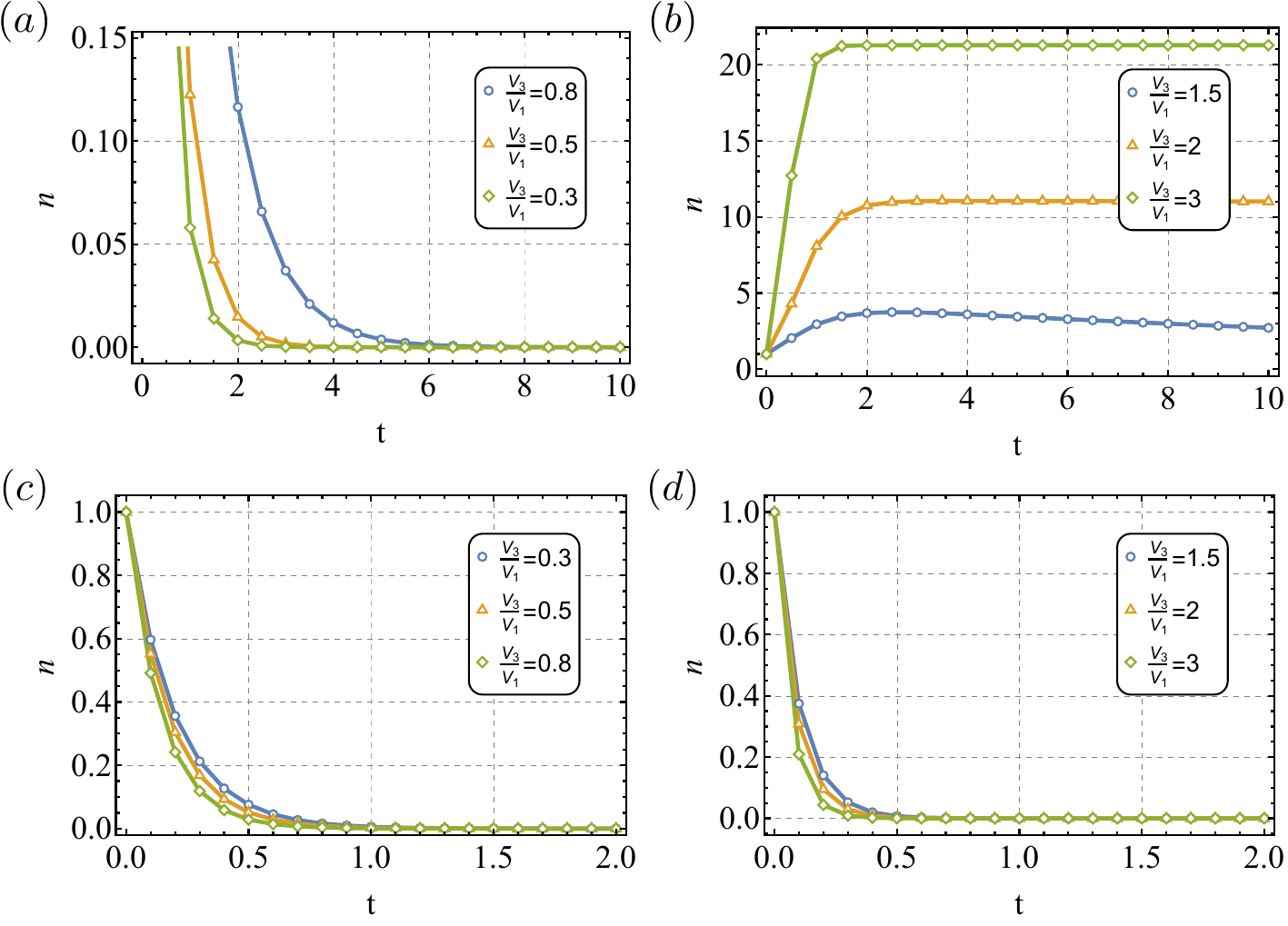}
    \caption{Numerical results of the evolution of average operator size $n$ over time $t$, obtained by solving \eqref{eqn:dP} for Model A under Definition I (a–b) and Definition II (c–d). Here, we set $N=100$. }
    \label{fig:num1}
  \end{figure}

\begin{figure}[tb]
    \centering
    \includegraphics[width=0.75\linewidth]{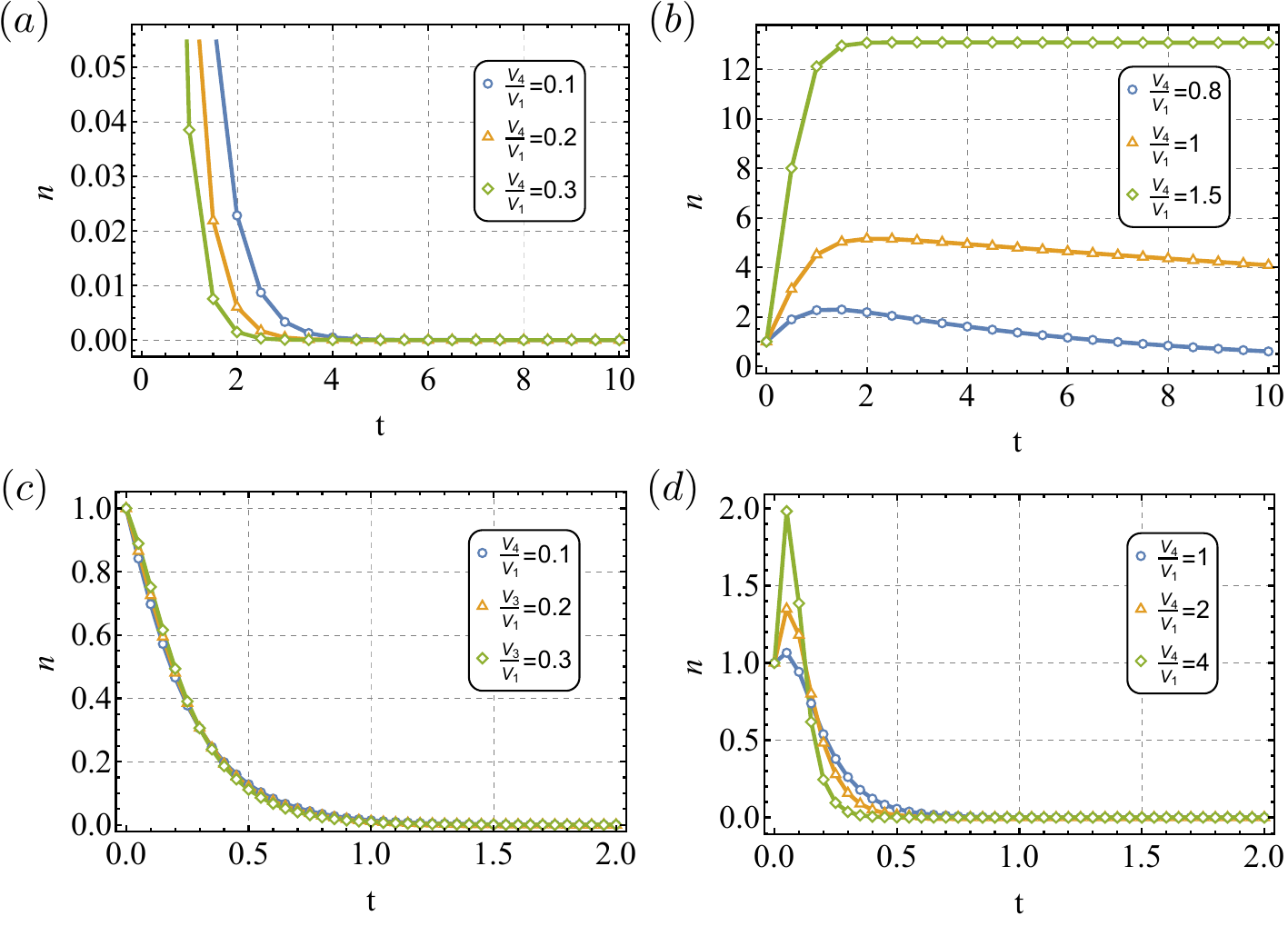}
    \caption{Numerical results of the evolution of average operator size $n$ over time $t$, obtained by solving \eqref{eqn:dP} for Model B under Definition I (a–b) and Definition II (c–d). Here, we set $N=100$. }
    \label{fig:num2}
  \end{figure}

The differential equation \eqref{eqn:dP} derived in this work is valid for arbitrary $N$, enabling a comprehensive investigation of finite-size effects. Motivated by results in the previous section, we first consider whether the operator size grows or decays in the short-time regime. Under either definitions, the short time behavior is determined by
\begin{equation}
\partial_t \overline{n}_{\text{I/II}}(0)=\sum_{n}n\partial_tP(n,0)=\sum_{n}n\sum_{\Delta n}C_{\Delta n}(n)\delta_{n+\Delta n,1}=\sum_n n~C_{1-n}(n).
\end{equation}
For Model A, straightforward calculation gives 
\begin{equation}
\partial_t \overline{n}_{\text{I}}(0)=-4V_1 + \dfrac{4V_3}{N^2}(N-1)(N-2).
\end{equation}
This suggests the existence of distinct dynamical behaviors separated by the ratio $r^*=\dfrac{V_1}{V_3} = \dfrac{(N-1)(N-2)}{N^2}$. When taking the limit of $N\rightarrow \infty$, we find $r^* \rightarrow 1$, which matches the critical point of the scrambling transition. This finite-$N$ shift of the critical point is not captured by calculations based on the scramblon effective theory. On the other hand, we find $\partial_t \overline{n}_{\text{II}}(0)<0$ for arbitrary $V_1/V_3>0$. For Model B, we find 
\begin{equation}
\partial_t\bar{n}_{\text{I}}(0) =\partial_t\bar{n}_{\text{II}}(0)  =-4V_1 + \dfrac{8V_4}{N^3}(N-1)(N-2)(N-3).
\end{equation}
Therefore, there is no difference between the two definitions of operator size in open systems during the early-time regime, even at finite $N$. These theoretical predictions match the numerical solution of the equation \eqref{eqn:dP}, as shown in Figure \ref{fig:num1} and \ref{fig:num2} for various $V_3/V_1$ or $V_4/V_1$ with $N=100$.   

\begin{figure}[tb]
    \centering
    \includegraphics[width=0.75\linewidth]{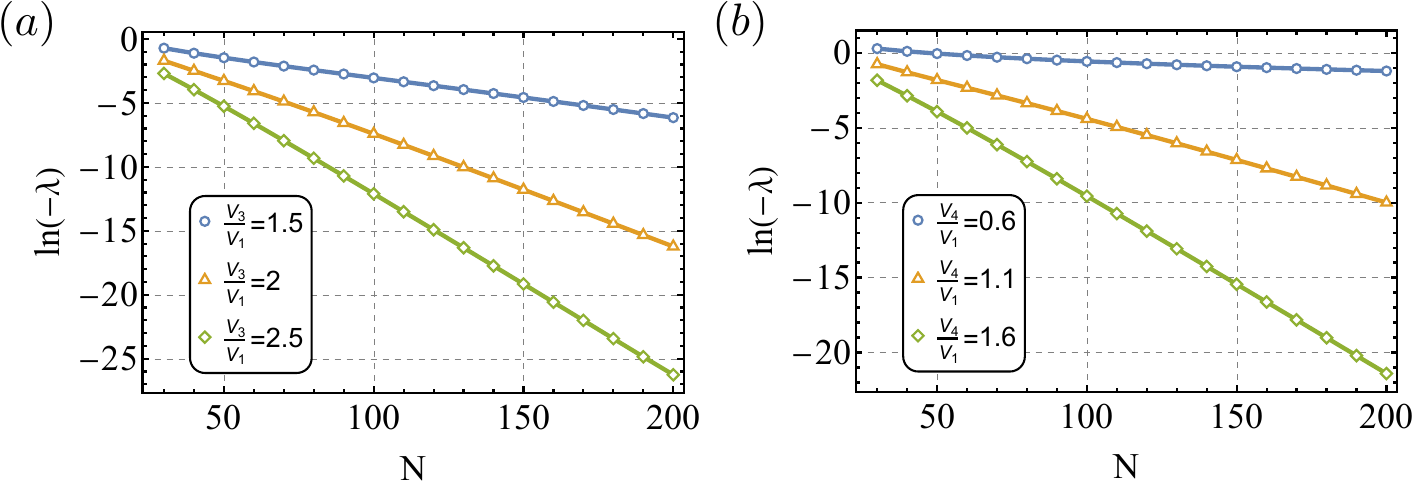}
    \caption{Numerical results for the second-largest eigenvalue $\lambda$ of the effective Lindblad operator $\mathcal{L}_s^{\text{I}}$ for Model $A$ (a) and Model B (b), respectively, clearly demonstrate the scaling behavior $\lambda \sim e^{-\# N}$. }
    \label{fig:num3}
  \end{figure}

However, a closer examination of the long-time behavior under Definition I reveals another interesting feature that is not captured by scramblon theory. While the scramblon theory predicts that the operator size saturates to an $O(N)$ value in the scrambling phase, numerical results instead show a gradual decay—albeit with a decay rate much smaller than that observed in the dissipative phase or under operator dynamics defined by Definition II. We study this decay rate by calculating the eigenvalue of $\mathcal{L}_s^{\text{I}}$ using the explicit matrix element \eqref{eqn:222} and \eqref{eqn:111}. Since the identity operator are decoupled from other operators, we always have a eigenvector $P(n,t)=\delta_{n0}$ with eigenvalue $0$. Therefore, the long-time decay of the the operator size is dominated by the second-largest eigenvalue, denoted as $\lambda$. The result is presented in Figure \ref{fig:num3}, showing an exponential dependence on the system size $N$. Intuitively, this is because even after saturating to the long-time distribution predicted by scramblon theory, there remains an exponentially small operator weight at small sizes $n\sim O(1)$. These small-size operators are still coupled to the $n=0$ sector, which leads to a slow decay. Such decay processes are neglected in the scramblon calculation.

\section{Summary}
In this work, we explore the dynamics of operator size in open quantum systems, addressing the challenge of reconciling two distinct definitions of operator size that arise from different treatments of the system's interaction with its environment. Using solvable Brownian SYK models, we derive the differential equations governing the evolution of the operator size distribution for both definitions. The results show that, under Definition I, the operator dynamics exhibit a scrambling transition for both intra-system interactions and system-bath couplings involving multiple system operators. In contrast, under Definition II, the signature of a scrambling transition appears only when intra-system interactions are present. Additionally, we extend previous analyses by investigating finite-size effects, uncovering early-time scrambling behavior and late-time decay processes that were not captured by scramblon effective theories. These findings advance the understanding of information scrambling and dissipation in open quantum systems, offering insights relevant to quantum many-body physics, quantum information theory, and the design of quantum technologies.

\section*{Acknowledgment}
This project is supported by the NSFC under grant 12374477, the Shanghai Rising-Star Program under grant number 24QA2700300, and the Innovation Program for Quantum Science and Technology 2024ZD0300101.

\bibliography{Draft.bbl}

\end{document}